\documentclass[mypaper,7pt,twoside]{CoAst}
\usepackage{epsf,graphicx,fancyhdr}
\renewcommand{\chaptermark}[1]{\markboth{#1}{}}

\newcommand{\teff}{\ensuremath{T_{eff}}}             

\newcommand{\kopf}{\small\itshape Comm. in Asteroseismology\\ Vol. 143, 2003}
\newcommand{\Authors}[1]{\begin{center}\normalsize\bf\sf #1 \end{center}}

\renewcommand{\author}[1]{\begin{center}\normalsize\bf\sf #1 \end{center}}
\newcommand{\Address}[1]{\begin{center}\small\sf #1 \end{center}}

\renewenvironment{abstract}{\section*{Abstract}\normalsize\sf}{}
\newcommand{\References}[1]{\begin{flushleft}{\large References\\}\vspace*{2mm}\small #1 \end{flushleft}}

\newcommand{\chapterDSSN}[2]{\chapter[\sf\normalsize #1\\ \footnotesize \hspace*{5mm}by #2 \sf\normalsize][]{#1\\}\rhead[\fancyplain{}{\sf\footnotesize \center{#1}}]{\fancyplain{}{\sffamily\thepage}}\lhead[\fancyplain{\kopf}{\sffamily\thepage}]{\fancyplain{\kopf}{\sf\footnotesize \center{#2}}}}

\renewcommand{\chaptername}{}
\newcommand{\acknowledgments}[1]{\vspace*{5mm}\noindent\begin{bf}Acknowledgments. \end{bf} #1}

\begin{document}
\sf

\chapterDSSN{Instability strips of main sequence B stars: a parametric study of iron enhancement}{A. Miglio, P.-O. Bourge, J. Montalb\'an, M.-A. Dupret}
\Authors{A. Miglio$^1$, P.-O. Bourge$^1$, J. Montalb\'an$^1$, M.-A. Dupret$^2$} 
\Address{$^1$ Institut d'Astrophysique, All\'ee du 6 Ao\^ut, 17, B-4000 Li\`ege, Belgium\\
$^2$ LESIA, Observatoire de Paris, F-92195 Meudon, France}

\noindent
\begin{abstract}
The discovery of $\beta$ Cephei stars in low metallicity environments, as well as the difficulty to theoretically explain the excitation of the pulsation modes observed in some $\beta$ Cephei and SPB stars, suggest that the iron opacity ``bump'' provided by standard models could be underestimated.
We investigate, by means of a parametric study, the effect of a local iron enhancement on the location of the $\beta$ Cephei and SPB instability strips.\\
\end{abstract}

The excitation of $\beta$ Cephei and SPB stars is challenging current theoretical models, since the latter cannot satisfactorily reproduce the observations of $\beta$ Cep stars in low metallicity environments (e.g. Ko{\l}aczkowski et al 2006),  the excitation of the observed pulsation modes in some $\beta$ Cep stars (Pamyathnyk et al. 2004, Ausseloos et al. 2004, Handler et al. 2005), the excitation of hybrid SPB and $\beta$ Cep pulsators, and the observation of SPB type pulsations in ``cool'' B-type stars (Antonello et al. 2006, Bruntt et al. 2006).

As discussed in Miglio et al. (2006), the current uncertainties on opacity calculations, and on the assumed metal mixture, have an significant impact on the excitation of modes in both $\beta$ Cep and SPB stars.  Nonetheless, these uncertainties may not be sufficient to explain all the discrepancy between theoretical predictions and observations.

\begin{figure}[!ht]
\centering
\includegraphics[clip,angle=0,width=0.73\textwidth]{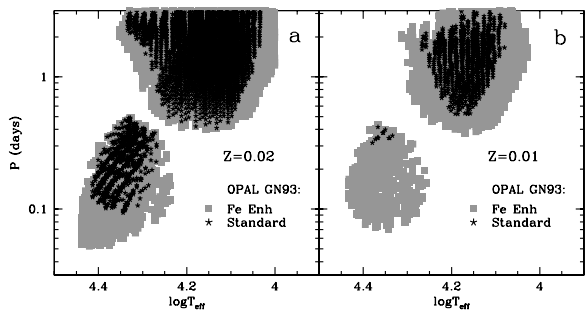}
\vspace{-0.8cm}
\caption{\small Instability strips represented in a $\log{\teff}$-$\log{P}$ diagram. In each panel, the two regions of unstable modes represent $\beta$ Cep- and SPB-type pulsations.}
\label{fig1}
\end{figure}

We therefore investigate another possible solution to the problem: we follow the suggestion by Cox et al. (1992) and Pamyatnykh et al. (2004) and carry out a parametric study of the effect of local iron enhancement on the stability of SPB and $\beta$ Cep stars. We base our parametric description on Fe accumulation profiles as found in models of A-F stars with diffusion and radiative accelerations (Richard et al. 2001; see also Bourge et al. 2006a,b for B stars).
At each time step in the evolution we increase the Fe mass fraction in the chemical mixture. 
The increase is described by a Gaussian function centred at $\log \mathrm{T} \sim 5.2$, where this value is justified by calculations of radiative accelerations using the \textsc{op} webserver (see also Bourge et al., these proceedings).
We calculate the opacity by interpolating between several \textsc{opal} tables computed with different mass fractions of Fe in the chemical mixture. 
Though the results depend on the Fe accumulation rate, we find (see Fig.~1) that the SPB/$\beta$ Cep instability strips get wider (in particular the SPB-type instability occurs also at lower $\mathrm{T_{eff}}$), higher frequency modes are excited in $\beta$ Cep models and a larger number of $\beta$ Cep models is found to be excited for Z=0.01.

\acknowledgments{A.M. and J.M. are supported by the PRODEX 8 COROT grant C90199, P.-O.B. by the Belgian IAP grant P5/36 and M.-A.D. by the CNRS. A.M. and J.M. also acknowledge E. Antonello and L. Mantegazza for their valuable collaboration and suggestions.}

\References{
Antonello E., Mantegazza, L., Rainer, M., Miglio, A. 2006, A\&A 445, L15\\
Ausseloos M., Scuflaire R., Thoul A., Aerts C. 2004, MNRAS 355, 352\\
Bourge P.-O., Alecian G. 2006a, ASP Conf. Ser. 349, 201\\
Bourge P.-O., Alecian G., Thoul A., Scuflaire R. Th\'eado S. 2006b, CoAst 147, 105\\
Bruntt, H., Southworth, J., Torres, G., Penny, A.~J., et al. 2006 A\&A 456, 651\\
{Cox} A. N., {Morgan} S. M., {Rogers} F. J., {Iglesias} C. A. 1992, ApJ 393, 272\\
Handler G., Jerzykiewicz, M., Rodriguez, E., et al. 2006, MNRAS 365, 327\\
Ko{\l}aczkowski Z., Pigulski, A., Soszy\'nski, I. et al. 2006, MemSait 77, 336\\
Miglio A., Montalb\'an J., Dupret M.-A. 2006, MNRAS, \emph{accepted}\\
Pamyatnykh A. A., Handler G., Dziembowski W. A. 2004, MNRAS 350, 1022\\
Richard O., Michaud G., Richer J. 2001, ApJ 558, 377\\
}

\end{document}